\documentclass{llncs}
\usepackage[utf8]{inputenc}
\usepackage{makeidx} 
\title{Trust Implications of DDoS Protection in Online Elections}
 
 \date{}

 \author{Chris Culnane\inst{1} \and Mark Eldridge\inst{2} \and Aleksander Essex\inst{3} \and Vanessa Teague\inst{4}}
 \institute{
 Department of Computer and Information Systems\\
 University of Melbourne\\
 \email{christopher.culnane@unimelb.edu.au}\and
 School of Computer Science\\
 University of Adelaide\\
 \email{mark.eldridge@student.adelaide.edu.au}\and
 Department of Electrical and Computer Engineering\\
 University of Western Ontario\\
 \email{aessex@uwo.ca}\and
 Department of Computing and Information Systems\\
 University of Melbourne\\
 \email{vjteague@unimelb.edu.au}
 }

\newcommand{\commentOutForArXiV}[1]{}

\commentOutForArXiV{
\usepackage{natbib}
\usepackage[numbers]{natbib}
\makeatletter
\renewcommand\bibsection%
{
\section*{\refname
\@mkboth{\MakeUppercase{\refname}}{\MakeUppercase{\
refname}}}
}
\makeatother
}
\usepackage{graphicx}
\graphicspath{ {images/} }
\usepackage[hyphens]{url}
\usepackage{hyperref}
\usepackage{listings}
\usepackage{xcolor}
\usepackage{keyval,xparse}
\usepackage[frozencache=true,outputdir=out,cachedir=cache]{minted}
\usepackage{xargs}   
\usepackage{color}
\usepackage{msc}
\usepackage{verbatim}
\usepackage{mdframed}
\usepackage[colorinlistoftodos,prependcaption,textsize=tiny]{todonotes}

\definecolor{shadecolor}{rgb}{0.95,0.95,0.92}
\definecolor{codegreen}{rgb}{0,0.6,0}
\definecolor{codegray}{rgb}{0.5,0.5,0.5}
\definecolor{codepurple}{rgb}{0.58,0,0.82}
\definecolor{backcolour}{rgb}{0.95,0.95,0.92}

\usepackage{rotating}
\usepackage[
  n,
  operators,
  advantage,
  sets,
  adversary,
  landau,
  probability,
  notions,  
  logic,
  ff,
  mm,
  primitives,
  events,
  complexity,
  asymptotics,
  keys]{cryptocode}

\def\mintedargs{linenos,breaklines=true,numbersep=5pt,fontsize=\footnotesize,bgcolor=shadecolor}

\newmintinline[javascript]{javascript}{bgcolor=shadecolor,breaklines}
\makeatletter

\define@key{code}{caption}{\def\mm@caption{#1}}
\define@key{code}{label}{\def\mm@label{#1}}
\define@key{code}{lang}{\def\mm@lang{#1}}
\define@key{code}{placement}{\def\mm@placement{#1}}
\DeclareDocumentCommand{\code}{O{htp} m m}{%
  \begingroup%
  \setkeys{code}{lang={},caption={CAPTION GOES HERE},label={lst:unknow},placement={ht arg},#2}%
  \expandafter\begin{listing}[#1]
  \expandafter\inputminted\expandafter[\mintedargs]{\mm@lang}{#3}\vspace{-8mm}\caption{\mm@caption}
    \label{\mm@label}
 \end{listing}
  \endgroup%
}

\makeatother

\begin{document}

\maketitle
\begin{comment}
\textit{
\section*{Things to Cover}
 \begin{itemize}
    \item Incapsula Certificate distribution - Analysis of IPs returning WA cert
    \item Incapsula Script injection - Uncontrolled injection of JS and ability to compromise PIN and VoterID - Proof of concept
    \item Incapsula same server for registration and voting - persistent cookie
    \item Brute force PIN/VoterID - Proof of concept
    \item MITM double encryption - partial votes
    \item Possibly not correctly checking Control Code return
    \item Decryption ceremony - VPN
    \item Incapsula's fingerprinting behaviour, how DDoS mitigation is performed (identifying valid users), and implications for targeting individual voters or groups using JavaScript injection
\end{itemize}
}
\end{comment}
\begin{abstract}

Online elections make a natural target for distributed denial of service attacks. Election agencies wary of disruptions to voting may procure DDoS protection services from a cloud provider. However, current DDoS detection and mitigation methods come at the cost of significantly increased trust in the cloud provider. In this paper we examine the security implications of denial-of-service prevention in the context of the 2017 state election in Western Australia, revealing a complex interaction between actors and infrastructure extending far beyond its borders.\\

Based on the publicly observable properties of this deployment, we outline several attack scenarios including one that could allow a nation state to acquire the credentials necessary to man-in-the-middle a foreign election in the context of an unrelated \emph{domestic} law enforcement or national security operation, and we argue that a fundamental tension currently exists between trust and availability in online elections.

\end{abstract}

%!TEX root = ../main.tex

\section{Introduction}

Democratically elected governments may still aspire to the old principle of being \emph{of the people, by the people, and for the people}. But when it comes to contemporary deployments of internet voting, the technology underpinning {\em how} governments are elected is a different story, and we are beginning to observe local elections carrying an increasingly multi-national footprint. 

In this paper we present an analysis of the 2017 state election of Western Australia (WA) as one such case study. We found a complex interaction between jurisdictions extending far beyond WA's borders. The election software was created by a Spanish based company. The election servers were hosted in the neighbouring state of New South Wales. Voters connected to the election website via a U.S. based cloud provider. They were presented with a TLS certificate that was shared with dozens of unrelated websites in countries such as the Philippines, Lithuania, and Argentina, and that was served out of data centers in countries such as Japan, Poland, and China.

In particular this work focuses on the implications of cloud-based distributed denial of service (DDoS) protection in an election setting, revealing the existence of a tension between availability and authentication.

\subsection{Background}

The acceptance of an election result should not come down to trust, but it often does.  Some systems, such as fully scrutinised manual counting, Risk Limiting Audits~\cite{lindeman2012gentle} and end-to-end verifiable cryptographic systems~\cite{Adi08,starvote,chaum2008scantegrity,ryan2009pret,rosenjonathan,kiayias2015demos}, allow voters and observers to derive evidence of an accurate election result, or to detect an inaccurate result.  

Australia's iVote Internet voting system, implemented by third-party vendor Scytl, does not provide a genuine protocol for verifying the accuracy of the election outcome, relying instead on a collection of trusted and semi-trusted authorities and auditors~\cite{halderman2015new}.  At the time of writing, it is the largest continuing Internet voting system in the world by number of votes cast.\footnote{The largest as a fraction of the electorate is Estonia's.}  The Western Australian run was, however, very small: about 2000 votes were received, out of an electorate of 1.6 million.  Election day was March 11th 2017, but iVote was available during the early voting period starting on 20th February.

For recent elections conducted in the Australian states of Western Australia and New South Wales, the iVote system was used in conjunction with Imperva Incapsula, a global content delivery network which provides mitigation of Distributed Denial of Service (DDoS) attacks.

DDoS attacks involve using a large number of connections to flood a target website, overloading systems and preventing legitimate users from logging in. It was a DDoS attack which was blamed for the failure of the Australian Government online eCensus system in August 2016~\cite{macgibbon_review_2016,2016census-issues-of-trust}. To mitigate these attacks, Incapsula's systems act as a TLS proxy, intercepting secure connections between the voter and the iVote servers and filtering malicious traffic.

Following our analysis of the unintended consequences of TLS proxying in the Western Australian Election, a subsequent by-election in New South Wales used Incapsula only for registrations and demonstration of iVote, not for the actual voting process itself. However, valid TLS certificates for the Western Australian and New South Wales election systems continue to be served by Incapsula servers all over the world. This illustrates the difficulty of reversing a decision to outsource trust.

\subsubsection{Contributions.}
Our contributions are threefold. Firstly, we provide an analysis of the front-end iVote protocol, including the associated credential exchange and key derivation. 

Secondly, we analyse the implications of running an internet voting system through a cloud based DDoS protection service acting as a non-transparent TLS proxy. We provide the results of a global scan to assess the scale with which Western Australian election related TLS certificates had been globally deployed. We identify and discuss the misconfigurations we discovered in the case of the Western Australian state election 2017, and analyse the feasibility of a malicious TLS proxy performing a brute force attack on voter credentials. 

Finally, we examine the injection of JavaScript performed by the DDoS protection service, and provide a proof of concept of how this could be utilised by a malicious entity to compromise voter credentials and modify ballots. We disclosed our findings to the Western Australian Electoral Commission, both before and during the election. They addressed the server misconfiguration, but continued to use the cloud based DDoS protection service for the duration of the election. 

%Analysis. Internet-wide scan. Disclosure.
%\todo{Need more text}

\subsubsection*{Paper Organization.}
The rest of the paper is organized as follows. Section~\ref{sec:ivote} describes the iVote protocol, and how a voter's cryptographic credentials can be recovered by a man-in-the-middle observing messages exchanged between the client and iVote server. Section~\ref{sec:ddos} describes technical findings of the cloud-based DDoS protection service, focusing on their certificate management practices. Based on these findings Section~\ref{sec:mitm} proposes two attack scenarios that could allow the cloud provider (or a coercive entity) to man-in-the-middle an election. Section~\ref{sec:additional} presents additional findings and Section~\ref{sec:conc} concludes.

% Section~\ref{sec:certs} describes where the valid certificates for the New South Wales and Western Australian elections have been found, and what the immediate implications are.  Section~\ref{sec:decryptableStored} explains why this allows Incapsula to decrypt votes, even without tampering with the client-side javascript it serves.  Section~\ref{sec:scriptInject} explains how easy it would be to tamper with client-side javascript in a way that would be very difficult to detect.  Section~\ref{sec:verifiability} explains why this would make iVote's very limited verification mechanism easier to subvert.  Finally, in Section~\ref{sec:bypassingDDoS}, we explain why the system was not properly protected from DDoS attacks until we notified the authorities.

%!TEX root = ../main.tex

\section{The iVote Protocol}
\label{sec:ivote}
In this section we describe the iVote protocol. In particular we observed that partial votes are sent---and stored on the server---encrypted by a symmetric key which is only protected by a key derived from the voter's ID and PIN. As we shall discuss, this leads to the potential to recover votes via a brute force attack of the \texttt{iVoteID} or \texttt{PIN}. When combined with the wider issue of using the same TLS Proxy for registration as voting, the brute force attack becomes viable. 

\subsection{Key Findings}

In iVote the secret keys used to construct an encrypted and digitally signed ballot are cryptographically derived from two values: a voter's ID and PIN. Knowledge of these two values is sufficient information to allow an attacker to impersonate a voter and cast a valid ballot on their behalf. iVote seemingly acknowledges the sensitivity of these values 

The key finding of this section is that the iterative hashing scheme used by iVote to protect the ID / PIN pair can be brute forced in practice by a man-in-the-middle observing messages exchanged between a voter's client and the iVote server. While transport layer security (TLS) protects these values from the view of most network observers, as we explain in Section~\ref{sec:ddos}, the non end-to-end nature of TLS in DDoS prevention exposes these values to the cloud provider.

\subsection{Methodology}
Publicly available technical documentation of the iVote system as deployed in WA is limited.  Significant information about the system and its configuration, however, can be observed from its public internet-facing components via a demonstration website set up by the Western Australian Electoral Commission (WAEC) to allow voters to practice voting. To test the implementation we created our own local server based on the publicly available JavaScript. There were, however, two main limitations to this approach: (1) the practice website did not include the registration step, and as such we were unable to observe network messages exchanged during this phase, and (2) the responses by the practice iVote server were mocked, and may not convey the full functionality of the live election website. Following our initial analysis, we contacted the WEAC on Feb 17th, 2017 with a report of our findings, which WAEC acknowledged the same day.

\subsection{Voter Experience}
An iVote election has three main phases:

\begin{enumerate}
\item {\bf Registration.} A voter visits a registration website, enters her name, her registered address and her date of birth. She may possibly be asked for further identifiers such as a passport number. She then chooses and submits a 6-digit PIN, which we will refer to as \texttt{PIN}.  An 8-digit iVote ID number, which we will refer to as \texttt{iVoteID}, is sent to her via an independent channel such as by post or SMS. 
\item {\bf Voting.} The voter visits the voting website and enters her \texttt{iVoteID} and her \texttt{PIN}.  Her vote is encrypted in her browser using JavaScript downloaded over TLS from the voting server.  If she wishes, she may pause voting and resume later---to facilitate this, a partially-completed vote is stored (encrypted) on the server while she is voting.  When she submits her vote, she receives a 12-digit receipt number.
\item {\bf Verification.} All submitted votes are copied to a third-party verification server.  After voting, the voter may call this service, enter her \texttt{iVoteID}, \texttt{PIN} and Receipt number, and then hear her vote read back to her.
\end{enumerate}

\subsection{Protocol Overview}

\begin{sidewaysfigure}
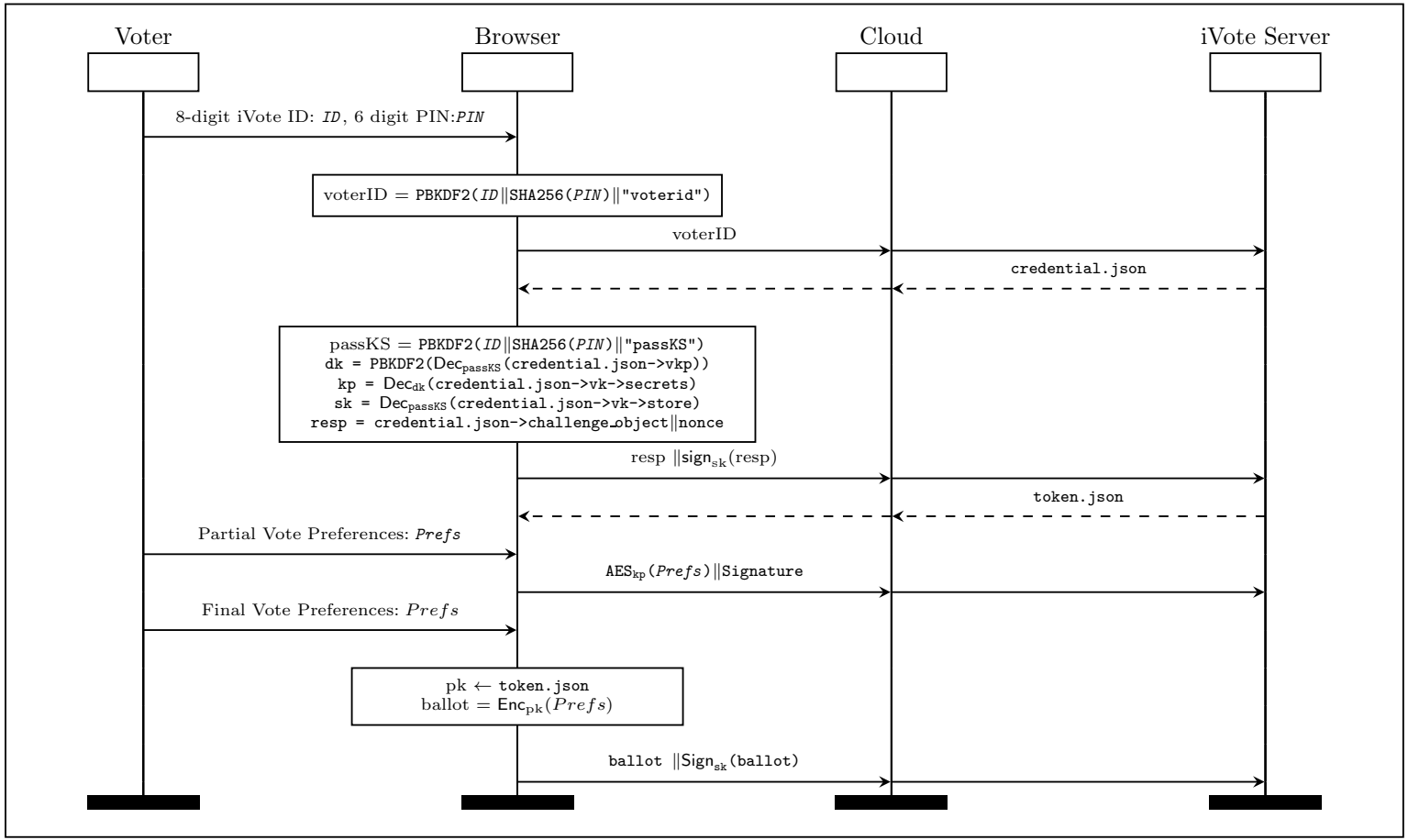

\centering
\begin{msc}[level height=0.55cm, msc keyword=]{}
\setlength{\instdist}{3.8cm}
\setlength{\firstlevelheight}{0.1cm}
\setlength{\lastlevelheight}{0.2cm}
\setlength{\topheaddist}{0.6cm}
\setlength{\topnamedist}{0.1cm}
\setlength{\topheaddist}{0.7cm}
\setlength{\bottomfootdist}{0.4cm}
\declinst{vtr}{Voter}{}
\declinst{vtrclient}{Browser}{}
\declinst{tlsprox}{Cloud}{}
\declinst{vcs}{iVote Server}{}
\nextlevel
\mess{{\scriptsize{8-digit iVote ID: \texttt{\textit{ID}}, 6 digit PIN:\texttt{\textit{PIN}}}}}{vtr}{vtrclient}
\nextlevel
\action*{\scriptsize{voterID = \texttt{PBKDF2(\textit{ID}$\|$SHA256(\textit{PIN})$\|$"voterid")}}}{vtrclient}
\nextlevel
\nextlevel
\mess{{\scriptsize voterID}}{vtrclient}{tlsprox}
\mess{}{tlsprox}{vcs}
\nextlevel
\mess*{\scriptsize{{\tt credential.json}}}{vcs}{tlsprox}
\mess*{}{tlsprox}{vtrclient}
\nextlevel
\action*{\parbox{6.6cm}{\centering\scriptsize{passKS = \texttt{PBKDF2(\textit{ID}$\|$SHA256(\textit{PIN})$\|$"passKS")\\ 
dk = PBKDF2($\mathsf{Dec}_{\text{passKS}}$({\tt credential.json->vkp}))\\
kp = $\mathsf{Dec}_{\text{dk}}$({\tt credential.json->vk->secrets})\\ sk = $\mathsf{Dec}_{\text{passKS}}$({\tt credential.json->vk->store}) \\ resp = {\tt credential.json->challenge\_object}$\|$nonce}}}}{vtrclient}
\nextlevel
\nextlevel
\nextlevel
\nextlevel
\mess{\scriptsize{resp $\| \mathsf{sign}_{\text{sk}}$(resp) }}{vtrclient}{tlsprox}
\mess{}{tlsprox}{vcs}
\nextlevel
\mess*{\scriptsize{{\tt token.json}}}{vcs}{tlsprox}
\mess*{}{tlsprox}{vtrclient}
\nextlevel
\mess{\scriptsize{Partial Vote Preferences: \texttt{\textit{Prefs}}}}{vtr}{vtrclient}
\nextlevel
\mess{\scriptsize{\texttt{AES$_{\text{kp}}$(\textit{Prefs})}$\|$\texttt{Signature}}}{vtrclient}{tlsprox}
\mess{}{tlsprox}{vcs}
% % Omitted for space
% \mess*{\scriptsize{\texttt{\{\}} \textit{empty response}}}{vcs}{tlsprox}
% \mess*{}{tlsprox}{vtrclient}
\nextlevel
\mess{\scriptsize{Final Vote Preferences: $Prefs$}}{vtr}{vtrclient}
\nextlevel
\action*{\parbox{4.5cm}{\centering\scriptsize pk $\leftarrow$ {\tt token.json} \\ ballot = $\mathsf{Enc}_{\text{pk}}(Prefs)$}}{vtrclient}
\nextlevel
\nextlevel
\nextlevel
\mess{\scriptsize{\texttt{ballot $\|\mathsf{Sign}_{\text{sk}}$(ballot)}}}{vtrclient}{tlsprox}
\mess{}{tlsprox}{vcs}

% % Omitted for space
% \nextlevel
% \mess*{\scriptsize{ConfirmationCode$\|$\texttt{Signature}}}{vcs}{tlsprox}
% \mess*{}{tlsprox}{vtrclient}
% \nextlevel
% \mess{\scriptsize{ReceiptNum$\|$ConfirmationCode}}{vtrclient}{vtr}
% \nextlevel

\end{msc}
\label{fig:ivote-protocol}
\caption{\textbf{iVote Protocol}. High-level overview of login and ballot casting protocol (\textit{n.b.,} some details omitted for brevity). The TLS connection is not end-to-end between the browser and iVote server, exposing brute-forcible voter credentials to the cloud provider.}
% \end{figure}
\end{sidewaysfigure}

A complete overview of the protocol is both beyond the scope of this paper, and beyond what can be observed from the public-facing elements of the system. We do, however, have sufficient information to outline how a brute force attack to recover voter credentials could proceed. A high-level overview of login and ballot casting is depicted in Figure~\ref{fig:ivote-protocol}, with additional details as follows.

\subsubsection{Login.}\label{sec:login} The voter first enters their \texttt{iVoteID} and \texttt{PIN} into the login page in the browser. A cryptographic key derivation implementation in client-side JavaScript then uses these values to derive a value, \texttt{voterID}, as follows. First a string is created of the form \texttt{iVoteID + "," + Base64(SHA256(PIN)) + "," + "voterid"}. This string is used as the password input to the password-based key derivation function \texttt{PKCS\#5 PBKDF2WithHmacSHA1}. The function uses the generated password along with a salt of 20 null bytes; it performs 8000 iterations and outputs a key of length 16 bytes. The result is hex encoded before being posted to the server as \texttt{voterID}. 

The purpose for this seems to be to protect the \texttt{iVoteID} and \texttt{PIN} by not sending them to the server directly.
However, as we discuss in Section~\ref{sec:bruteForce}, this protection is insufficient as it is computationally feasible to recover these values from \texttt{voterID} through brute-force search.

\footnotetext{This is an instance of a \texttt{PKCS\#5 PBKDF2WithHmacSHA1} function, with a salt consisting of 20 null bytes, performing 8000 iterations, and generating a key of 16 bytes}

\subsubsection{Voter Credentials.}\label{sec:credentialfile}
If the \texttt{voterID} submitted to the server corresponds to a registered voter, the server responds with a file {\tt credential.json}. An outline of this file is shown in Listing~\ref{lst:cred_skel}. The demo system uses an internal mocked response for a sample user, however we conjecture the real election server simply stores a database of \texttt{voterId}/\texttt{credential.json} pairs, and responds with the associated \texttt{credential.json} whenever a valid \texttt{voterID} is presented.

\code{lang=JavaScript, caption=Voter Credential File Skeleton, label=lst:cred_skel}{code/cred_skeleton.json}
%\code{lang=JavaScript, caption=vk Object Skeleton, label=lst:vk_skel}{code/vk_skeleton.json}

\noindent The \texttt{vad} object contains a number of keys and certificates. The \texttt{vk} object represents a Scytl KeyStore, which combines a \texttt{PKCS\#12} keystore with a JSON object of encrypted secrets. The underlying  \texttt{PKCS\#12} keystore is protected by what the code refers to as the  {\em long password}. The first step to deriving the {\em long password} is to derive an AES key to decrypt the password contained in \texttt{vkp}. To do this a string is created similar to the one created during the login phase. This string has the form:  \texttt{iVoteID + "," + Base64(SHA256(PIN)) + "," + "passKS"}. The string differs from the one constructed at login time using the suffix ``passKS" instead of ``voterid".

This password string, along with the salt value in \texttt{vkp}, is passed to another instance of \texttt{PKCS\#5 PBKDF2WithHmacSHA1} that performs 8000 iterations. The result is a 16-byte key, which is then used to initialise an AES cipher, using GCM (Galois/Counter Mode) with no padding. The GCM nonce length is 12 bytes and the tag length is 16 bytes. The nonce is the first 12 bytes of the password value stored in \texttt{vkp}. The remaining value of \texttt{vkp} is decrypted to form what the code calls the {\em derived password}.

The  {\em long password} is finally generated by a \texttt{PBKDF2WithHmacSHA1} that performs a single iteration on the {\em derived password} along with the salt value from \texttt{vk}, yielding a 16 byte key. This value is used as both a password for the \texttt{PKCS\#12} key store, and as an AES Key to decrypt the values in the secrets object. The keys in the {\tt secrets} object in \texttt{vk} are Base64Encoded ciphertexts. The {\em long password} is used to initialize an AES Cipher using GCM with no padding. The GCM nonce length is 12 bytes and the tag length is 16 bytes. The nonce is the first 12 bytes of the value in the {\tt secrets} object, with the remainder being the ciphertext that is to be decrypted.

The final outcome of this intricate sequence of client-side key derivations and decryptions is an AES symmetric key {\tt kp} which is used by the browser to encrypt partial votes, which we will continue in further detail in Section \ref{sec:partial_vote}. 

%The certificates stored in the ``vad" object are as follows:

%\begin{description}
%\item[eeca] Election Event CA 
%\item[svca] Services CA
%\item[azca] Authorities CA
%\item[ab] Used to verify XML Signatures in EML (Election Markup Language) that defines the candidates and races in the %election. 
%\end{description}

%With the exception of the ab certificate, which is used to verify XML signatures, a search of the JavaScript indicates that none of the other certificates are even read from the credential JSON. By examining the certificates we can see that they were created in September 2016 and expire in July 2017, The svca and azca certificates are signed by the eeca.

\subsubsection{Token.}
The {\tt credential.json} file is further processed and the contents extracted, in addition the server's signature on the received challenge is verified. In response to a valid signature, the browser generates a random nonce, concatenates it with the server's challenge, and returns this as a signed message. The purpose of this check appears to be a means of confirming that a client has successfully recovered their private signing key in the keystore.

%let's not mention either of the methods - they are a formatting problem and don't add anything. Can reword to remove
% go for it

% The justification for this, however, is unclear since the ability to recover the private signing key is based on knowing the \texttt{iVoteID} and \texttt{PIN}, which the client must know in order to have generated a valid {\tt voterID} value during the login phase. Moreover, even if a malicious client was replaying that message it would not be able to submit valid vote information or decrypt partial votes without knowing the \texttt{iVoteID} and \texttt{PIN}. 

%\code{lang=JavaScript, caption=Challenge Response Object, label=lst:chall_resp_obj}{code/challresp.json}

The response is posted to \url{vote-encoder/token/{voterKeysId}?v=1} where {\tt v} is taken from the configuration file, and {\tt voterKeysId} comes from the voter certificate common name, which contains the string ``VoterAuth\_" followed by the {\tt votersKeysId}. The {\tt voterKeysId} value is important because it is used during all subsequent posts, including voting and partial votes. It is unclear how this value is derived or who generates it, but we suspect it is generated by the registration server during the credential file generation.

Finally, the server responds to the token post with {\tt token.json} that contains the public-key group parameters for encrypting ballot preferences, the Election Markup Language for the various races, and any partial votes that have been recorded. The specifics of the encryption and signature of voter ballot preferences are outside the scope of this paper.

% We will discuss the partial vote value in more detail in Section \ref{sec:partial_vote}.

%\code{lang=JavaScript, caption=Token Response Skeleton, label=lst:token_resp_skel}{code/tokenresp.json}

%The portion of Listing \ref{lst:token_resp_skel} that is of most of interest to us is the ``eo" field which contains the encrypted partial vote if one has been submitted, which we will discuss further in Section \ref{sec:partial_vote}.

\subsection{Partial Votes}\label{sec:partial_vote}
When a voter completes a voting screen, either the Legislative Assembly (lower house) or Legislative Council (upper house), a partial vote of all currently entered preferences is created and sent to the server. The submission is sent to \url{vote-encoder/partial_vote/{voterKeysId}?v=1}, with JSON object shown in Listing~\ref{lst:partial_vote_skel}.
\code{lang=JavaScript, caption=Partial Vote Skeleton, label=lst:partial_vote_skel}{code/partialvote.json}
The \texttt{eo} string is encrypted with the secret key contained in the \texttt{secrets} object in {\tt credential.json}, which was extracted as part of the credential file processing, discussed in the previous section. When a partial vote is contained within the Token Response the same AES key contained in the \texttt{secrets} object is used to decrypt its contents and restore the screen for the user. The crucial consequence of this is that unlike the final vote which is submitted under the encryption of a randomly generated AES key, which is in turn encrypted with the public key of the election, the partial vote is only protected by the AES key stored in the credential file. 

Given that the credential file itself is only protected by the an encryption key derived from the \texttt{iVoteID} and \texttt{PIN}, if the \texttt{iVoteID} and \texttt{PIN} are susceptible to brute force attacks, both the receiving server, and any TLS proxies in between, would have the ability to recover votes. The attack is not mitigated by the fact the final vote could be different, since the partial votes are always submitted as the voter moves between the screens, and as such, the attacker need only look for the last partial vote submission prior to final submission to be sure of the contents of the final vote. 

% % Aleks: not sure this level of detail is needed...
% \begin{verbatim}
% SG1798-CWO-01,SG1798_Lowbell_LA__JONES_Grace|1~SG1798_Lowbell_LA...
% \end{verbatim}                            
%__BROWN_Norman|2~SG1798_Lowbell_LA__WILSON_Sam|3:SG1798-09-01,atl,F|1::

\subsection{Brute Forcing Voter Credentials}
\label{sec:bruteForce}

One important question is how hard it would be for a man-in-the-middle to recover a voter's credentials from observed messages exchanged between the browser and iVote server. Since WA opted to disable re-voting for their election, a near real-time attack capability is needed in order to construct a valid (but malicious) ballot and transparently swap it into the voter's session before they can cast. We now show that this requirement can feasibly be satisfied in practice.

As described in Section~\ref{sec:login} the {\tt voterId} value sent by the browser at login time is derived from the voter's $iVoteID$ and $PIN$, and knowledge of these values would be sufficient to recover all the voter's other cryptographic values from {\tt credential.json} and {\tt token.json} files.

Recall the {\tt voterID} value is essentially 8000 iterations of {\tt SHA1} applied to $iVoteID$, an 8-digit system-assigned value concatenated with $PIN$, a 6-digit user-chosen value. This implies a brute-force upper bound of 
\[8\cdot 10^3 \cdot 10^8 \cdot 10^6 \approx 2^{60}\]
\noindent operations. In other words, the {\tt voterID} value provides 60 bits of security in the best case.

This falls well below the minimum recommended NIST 112-bit security level~\cite{nistkeys}.  As a comparison, at the time of writing the Bitcoin network was able to perform $2^{62}$ {\tt SHA1} hashes per second.\footnote{\url{https://blockchain.info/stats}} 

In practice, however, the voterID space may not be uniformly distributed. Only a few thousand $iVoteID$s were actually used.  Moreover since the registration server is also covered by the DDoS cloud provider, we may assume that a man-in-the-middle would also be able to observe the set of $iVoteID$s in the context of the registration step and associate an ID with a unique IP address. Under the assumption of a known $iVoteID$, the search space to recover the voter's credential would be
\[8\cdot 10^3 \cdot 10^6 \approx 2^{33}\]
\noindent hashes. This space could be searched nearly instantly using a moderately sized GPU cluster. For example, contempoary bitcoin mining ASICs now achieve hash rates in the tera-hash-per-second ({\it i.e.}, $>2^{40}$) range. Investment in expensive and difficult to procure custom hardware, however, is not necessary. The rise of inexpensive elastic cloud computing puts this attack within reach of nearly any budget, and recent work has examined offering crypto brute forcing as a service. Heninger {\it et al.}~\cite{faas}, for example, have deployed hundreds of concurrent instances on Amazon EC2 in pursuit of factoring RSA moduli.

As a more immediate timing comparison demonstrating the real-world feasbility of this attack, we implemented our own program to brute force {\tt voterIDs} in a threaded Python program using the Hashlib implementation of PBKDF2 and deployed it on Digital Ocean. Using a single  20-core droplet, our unoptimized (non-GPU) implementation was able to recover a 6-digit PIN in approximately 7 minutes at a cost of USD~\$0.11. With 10 concurrent droplets (Digital Ocean's default account max) the time to recovery is less than 1 minute, which we believe would plausibly be less than the time taken by the average vote to read, mark and cast a ballot. Using a GPU-optimized hashing implementation (e.g., Hashcat), however, we expect this time can be reduced to the millisecond range while retaining a comparable cost of pennies per recovered credential.

%!TEX root = ../main.tex

\section{Distributed Denial of Service Protection} \label{sec:ddos}

Imperva Incapsula is a US-based cloud application delivery company which provides numerous security services to websites including prevention and mitigation of DDoS attacks. In this section we present a technical analysis of relevant aspects of their service as used by the Western Australian Electoral Commission (WAEC) for the 2017 WA State Election. 

\subsection{Key Findings}

Our key finding in regards to the DDoS prevention service deployed in the 2017 WA State Election are threefold: 

\begin{enumerate}
    \item Encryption is not end-to-end between the voter and the iVote server;
    \item The cloud provider's practice involves the bundling of dozens of unrelated website domains into a single certificate's subject alternate name (SAN) list; and
    \item An internet-wide scan we conducted found valid TLS certificates for the election website being served by servers around the world.
\end{enumerate}

\noindent Taken together we argue that this opens the possibility of a foreign nation being able to obtain the private key necessary to man-in-the-middle WA voters through an unrelated domestic law enforcement or national security operation.  It also risks compromising the election as a result of error or malfeasance by server administrators all over the world.

Additionally, we discovered that the system initially deployed for the election did not correctly protect against DDoS attacks, despite the presence of Incapsula's DDoS mitigation service. Due to misconfiguration of the iVote server, we were able to determine the true IP address for the WA iVote server via historical domain registrations for the NSW iVote system used in 2015, which was also being used to host the WA iVote system. 

Upon discovering this vulnerability we notified the WAEC, who reconfigured the server to stop accepting connections that did not originate from Incapsula's systems.  

% % AE: Do we know this for a fact...?
% It fortunate that no DDoS attempt was made before the issue was resolved.

\subsection{Non End-to-End TLS}

In a typical TLS handshake the server presents its certificate to the client. Completing a TLS handshake takes time, and saving the session state requires the server allocate memory. This and other strategies allow attackers with with access to numerous hosts to overwhelm a server by flooding it with connection requests. When a DDoS mitigation service is involved, the TLS handshake is slightly altered to allow the service to identify and filter malicious requests by forcing incoming connections to be made through its infrastructure before being forwarded on to the destination in a separate connection. The result is that the service provider becomes an intermediary for all traffic to the iVote server.

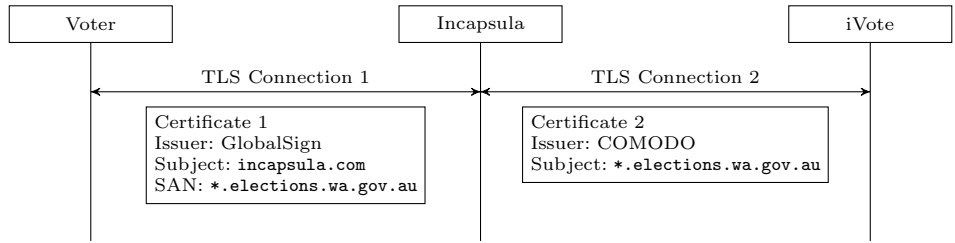
\begin{figure*}[t!]
	\centering
	\begin{tikzpicture}[node distance=3cm,auto,>=stealth']
	  % Initial
	  \tikzstyle{every node}=[font=\scriptsize]
	    \node[style={rectangle, draw=black, text width=6em,align=center, minimum height=1.5em}] (server) {iVote};
	    \node[left = of server, style={rectangle, draw=black,  text width=6em,align=center, minimum height=1.5em}] (mitm) {Incapsula};
	    \node[left = of mitm,style={rectangle, draw=black,  text width=6em,align=center, minimum height=1.5em}] (client) {Voter};
	    \node[below of=server, node distance=3cm] (server_ground) {};
	    \node[below of=mitm, node distance=3cm] (mitm_ground) {};
	    \node[below of=client, node distance=3cm] (client_ground) {};
	    \draw (client) -- (client_ground);
	    \draw (mitm) -- (mitm_ground);
	    \draw (server) -- (server_ground);

	    \draw[<->] ($(client)!0.30!(client_ground)$) -- node[above,scale=1,midway]{TLS Connection 1} node[draw,align=left, below=0.6em,scale=1,midway]{Certificate 1\\ Issuer: GlobalSign \\ Subject: {\tt incapsula.com}\\SAN: {\tt *.elections.wa.gov.au}} ($(mitm)!0.30!(mitm_ground)$);
	    
	    \draw[<->] ($(mitm)!0.30!(mitm_ground)$) -- node[above,scale=1,midway]{TLS Connection 2} node[draw, align=left,below=0.6em,scale=1,midway]{Certificate 2\\ Issuer: COMODO\\Subject: {\tt *.elections.wa.gov.au}} ($(server)!0.30!(server_ground)$);
	\end{tikzpicture}
	\caption{\textbf{Non end-to-end TLS}. Communication between a voter's browser and the iVote server pass through an Incapsula server and are decrypted, inspected, and re-encrypted under a different key.}
	\label{fig:non-e2e}
\end{figure*}

Incapsula's DDoS mitigation service operates by placing Incapsula servers between the user and the destination website as a non-transparent TLS proxy, intercepting all communications to and from the website in order to filter malicious connections. For example, when connecting to the iVote Core Voting System (CVS) at \url{https://ivote-cvs.elections.wa.gov.au}, the voter's connection first travels to a server owned by Incapsula where it is decrypted, scanned, and then forwarded on to the iVote server managed by the WAEC. This interaction is shown in Figure~\ref{fig:non-e2e}.

Nominally, if the iVote server was correctly covered by DDoS prevention, we should not have been able to observe its certificate, as the server would ignore any connection originating from a non-Incapsula IP address\footnote{\url{https://www.incapsula.com/blog/make-website-invisible-direct-to-origin-ddos-attacks.html}}. However, a misconfiguration of the iVote server made it possible to identify its true IP address, allowing us to request its TLS certificate directly. This issue is discussed in more detail in Section~\ref{sec:bypassingDDoS}.

The interception of connections allows Incapsula to filter out malicious traffic during DDoS attacks, but also allows Incapsula to see all traffic travelling through their systems. This behaviour is by design: modern DDoS mitigation methods rely on scanning the plaintext traffic being transmitted to the server they are protecting~\cite{incapsula_bits_bytes,cloudflare_ssl_faq}. Without this ability, they would have a much harder time determining the good connections from the bad ones. What it means, however, is that the voter's interaction with the voting server exists as plaintext at some point after leaving the voter's computer, but before reaching the election servers.

This fact is problematic since TLS authentication remains the only meaningful form of server authentication in iVote, and using a cloud provider for DDoS protection necessarily outsources this trust. Putting valid keys on a variety of third-party servers throughout the world brings all of them into the set of trusted parties, and increases the likelihood of a key leaking. Furthermore, ballot secrecy in iVote depends critically on the assumption that a voter's identity disclosed during registration cannot be linked with a cast ballot making non end-to-end encryption a concern in this matter as well.

\subsection{Large-scale Certificate Sharing}\label{large-scale-cert-sharing}

\begin{figure}[t!]
\begin{mdframed}
\scriptsize\texttt{incapsula.com, *.1strongteam.com, *.absolutewatches.com.au, *.advancemotors.com.au, *.alconchirurgia.pl, *.amplex.com.au, *.bohemiocollection.com.au, *.cheapcaribbean.com, *.compareit4me.com, *.elections.wa.gov.au, *.everafterhigh.com, *.farmerslifeonline.com, *.floraandfauna.com.au, *.heypennyfabrics.com.au, *.homeaway.com.ph, *.jetblackespresso.com.au, *.lifemapco.com, *.lovemyearth.net, *.maklernetz.at, *.mobile-vertriebe.de, *.mobile.zurich.com.ar, *.monsterhigh.com, *.mycommunitystarter.co.uk, *.noosacivicshopping.com.au, *.oilsforlifeaustralia.com.au, *.planetparts.com.au, *.purina.lt, *.redsimaging.com.au, *.rlicorp.com, *.roundup.fr, *.sassykat.com.au, *.spendwellhealth.com, *.sublimation.com.au, *.uat.user.zurichpartnerzone.com, *.woodgrove.com.au, *.yamahamotor-webservice.com, *.zlaponline.com, *.zurich-personal.co.uk, *.zurich.ae, *.zurich.co.jp, *.zurich.es, *.zurich.jp, *.zurichlife.co.jp, *.zurichseguros.pt, 1strongteam.com, absolutewatches.com.au, advancemotors.com.au, alconchirurgia.pl, amplex.com.au, bohemiocollection.com.au, compareit4me.com, farmerslifeonline.com, floraandfauna.com.au, heypennyfabrics.com.au, homeaway.com.ph, jetblackespresso.com.au, lifemapco.com, lovemyearth.net, mycommunitystarter.co.uk, noosacivicshopping.com.au, oilsforlifeaustralia.com.au, planetparts.com.au, purina.lt, redsimaging.com.au, roundup.fr, sassykat.com.au, spendwellhealth.com, sublimation.com.au, woodgrove.com.au, zurich.ae, zurich.es, zurich.jp, zurichlife.co.jp}
\end{mdframed}

\caption{{\bf Subject alternate names in the Incapsula certificate}. The same digital certificate used to prove the identity of {\tt *.elections.wa.gov.au} to WA voters is also used to prove the identity of websites listed above. This list was transient and changed several times in the month leading up to election day.}
\label{fig:hk-certificate}
\end{figure}

DDoS protection need not require a customer to surrender its private keys to the cloud provider~\cite{incapsula_bits_bytes,cloudflare_ssl_faq}. Instead, Incapsula outwardly presents their own certificate in the handshake, which includes the iVote server's domain (\url{ivote-cvs.elections.wa.gov.au}) in the Subject Alternate Name (SAN) extension of their certificate. Specifically Incapsula includes the wildcard domain \path{*.elections.wa.gov.au} in the SAN.

Obtaining this secondary certificate is a financial expense, and  Incapsula shares one certificate among numerous websites in order to reduce cost~\cite{incapsula_bits_bytes}. Specifically it lists itself as the certificate's subject, and packs numerous domains of its customers' into a single certificate's SAN. When a WA voter visits the iVote website \path{https://ivote-cvs.elections.wa.gov.au}, their browser is presented with a certificate with dozens of other  unrelated domains in the SAN. A list of these domains is given in Figure~\ref{fig:hk-certificate}, and includes websites for widely varying sectors and countries of origin. 

Through a combination of collecting our own TLS handshakes with the iVote server as well as Censys\cite{censys15} data we observed this certificate over a two month period prior to the election and found the SAN list changed several times, presumably as some clients joined and others left. For example, on Feb 1st the SAN included several casinos (\path{pandora-online-casino.com, }\path{caribiccasino.com}, \path{regalo-casino.com},\path{doublestarcasino.com}), but they disappeared shortly after. Importantly, visitors to any of these other websites are, in turn, presented with the {\em same} certificate.

% The actual server hosting this certificate (\url{107.154.128.220}) is located in an Incapsula Point of Presence (PoP) in Melbourne, Australia.

% This behaviour is seen by any user connecting to the iVote CVS domain from within Australia, due to the design of Incapsula's systems: users who perform a DNS lookup on any domain protected by Incapsula's services are directed to their nearest PoP, rather than the actual server.

\subsection{International Certificate Footprint}\label{international-cert-footprint}

Incapsula's global network consists of 32 data centres (Points of Presence, or PoPs), located across the Americas, Europe, the Middle East, and the Asia Pacific region.\footnote{\label{incapglobal}\url{https://www.incapsula.com/incapsula-global-network-map.html}} Due to the design of Incapsula's network, TLS certificates hosted in one PoP are propagated worldwide, so that users in any region served by Incapsula can have their connection proxied by the nearest PoP available. As stated by Incapsula:\footnote{\url{https://www.incapsula.com/blog/incapsula-ssl-support-features.html}} \textit{``When using Incapsula, our servers become the intermediate for all traffic to your website, including SSL traffic. To facilitate this, Incapsula needs a valid SSL certificate for your domain installed on all its servers worldwide.''}

We found Incapsula servers serving valid TLS certificates for \url{*.elections.wa.gov.au} from locations around the world, including Eastern and Western Europe, China, North and South America, and various points in Australia. 

These servers were identified through domain name look-ups for \url{ivote-cvs.elections.wa.gov.au} originating from within each country, and subsequent TLS connections, using a Virtual Private Network (VPN). Our timing analysis strongly indicates that the TLS certificates were being served directly by these servers, and not proxied from elsewhere.

\subsubsection{Internet Scan.}

We conducted an internet wide scan of the IPv4 space on election day (March 11, 2017), collecting all TLS certificates served over port 443 using {\tt zgrab}.\footnote{\url{https://github.com/zmap/zgrab}} In total we found 153 distinct IPs serving certificates containing \path{*.elections.wa.gov.au} in the subject alternate name. A traceroute and timing analysis showed that these IPs were consistent with cities in which Incapsula advertises data centers.\textsuperscript{\ref{incapglobal}} We were able to identify points of presence serving WA's certificate in Australia, Canada, China, France, Germany, Japan, Poland, Singapore, Spain, Switzerland, United Kingdom, and throughout the United States.

\section{Man in the Middle Attack Scenarios}
\label{sec:mitm}

In this section we outline two scenarios in which a man-in-the-middle could recover credentials necessary to be able to cast a valid ballot on a voter's behalf. 

\subsection{Modify the Scripts the DDoS Provider is Already Injecting} \label{sec:scriptInject}

\subsubsection{Overview and Significance.} In this first scenario, a malicious cloud provider injects Javascript into the voter's client with the aim of capturing their credentials. Since the cloud provider sits between the voter and iVote server, injecting a malicious script is an obvious but risky approach for the cloud provider if both the presence the script and its malicious purpose were detected. The significance of our particular attack scenario, however, makes use of the following observations: (1) the cloud provider is already rewriting server content to injecting their own JavaScript as part of their DDoS profiling functionality, and (2) the script payloads are already being obfuscated. 

We created a proof-of-concept vote-stealing script that leaks the voter's ID and PIN in the tracking cookie, and incorporated it into the script already being injected by the cloud provider at no increased file size.

\subsubsection{Script Injection for System Profiling.} 
When a voter connects to the iVote WA Core Voting System using the address \url{https://ivote-cvs.elections.wa.gov.au}, the connection is proxied through Incapsula's servers using an Incapsula-controlled TLS certificate. The initial response to a voter's connection sets a number of Incapsula cookies. 
% % AE: Pruned for space
% One is a persistent cookie, prefixed with the name ``visid\_incap\_" which expires approximately 1 year after creation. The second cookie is a session cookie, prefixed with ``incap\_ses\_". Both cookies contain Base64 encoded values that are different to each other. The exact contents is not known, since there appears to be some further encoding/encryption on the underlying value. We determined these values differ across independent connections, i.e. a clean machine with no cookies will get a different value. 

% These cookies appear to be used as part of the Incapsula DoS protection, as described by Incapsula in their blog post: \emph{How Incapsula Protects Against Data Leaks}~\cite{incapsula_protect_leaks}. These cookies act as a tracking cookie across all sub-domains of \path{*.elections.wa.gov.au}. As a result the same tracking cookie covers both the voting service and the registration service, allowing voters to be tracked across the two services if they use the same browser for both iVote registration and voting.

In addition the response is modified by Incapsula to include JavaScript code at the end of the HTML response. The included code inserts a \texttt{<script>} element to cause the browser to load an additional JavaScript file, the contents of which are obfuscated as a string of hex values. The included code is designed to perform fingerprinting of the voter's system. The HTTP responses for the resource files do not contain {\sf x-cdn} or {\sf x-iinfo} headers, strongly suggesting they are served by the Incapsula proxy (as would be expected), rather than by the iVote server.

When expanded into a more readable format, the injected JavaScript code is revealed as a tracking function. The code is designed to probe various parts of the voter's computer, including: the web browser they are using; any browser plugins they have installed; the operating system being used; their CPU type; and other information designed to fingerprint individual user connections. Additionally, this cookie calculates a digest of all other cookies set on the page, including those set by the server. 

This information is written into a profile cookie that is temporarily stored on the voter's computer. This profile cookie has an extremely short life of just 20 seconds, after which it will be deleted. Due to this being loaded during the page load the remaining requests within the page will send this cookie to the server before it disappears from the voter machine. As such, unless spotted within the 20 second period, or all requests/responses are being logged by the voter, it will be difficult for a voter to detect that this profiling cookie was ever set or sent to the server. The cookie is named {\tt \_\_utmvc}, which is similar to a Google Analytics cookie ({\tt \_\_utmv}), however, it does not appear to be related. The Google {\tt \_\_utmv} cookie is a persistent cookie used to store custom variables. The reason for the choice of naming is not immediately clear.

%\code{lang=html, caption=The voter information read by the obfuscated JavaScript code, label=lst:de-obfuscated-objects}{code/de-obfuscated-objects.js}

\subsubsection{Cookies and Voting}

While the concept of profiling and tracking cookies may seem invasive, there is nothing overtly malicious about this behaviour. Indeed, the entire web advertising industry is built to perform similar tasks, in order to track individual users across websites and better serve advertisements. 

For Incapsula, the tracking cookie most likely forms part of the DDoS mitigation process: Incapsula can determine which requests are likely to be from legitimate users. Combined with the profiling cookie, Incapsula can perform an analysis of the requesting device and alter its behaviour accordingly. 

%\subsubsection{JavaScript Injection and Security}

In the context of iVote, however, this behaviour poses a significant risk for voter security. As discussed in the introduction to this article, the iVote system is designed with the assumption that the encryption and authentication covering the communication between voter and server (Transport Layer Security, or TLS) is secure. If a third party has the ability to intercept this communication and inject malicious JavaScript into server responses, it would be possible to hijack the entire voting process. 

%Furthermore, the process for approving the voting system will involve the WA Electoral Commission signing off on the system---if the production system includes additional JavaScript which is not under the control or purview of the Electoral Commission, it would seem to breach this oversight process. 

The JavaScript we have witnessed being injected into server responses is non-malicious, however, there remains the potential for this to not always be the case. For example, a rogue Incapsula employee or a foreign intelligence service with access to Incapsula's systems could alter the injected JavaScript. If this occurred, it would be possible to steal the \texttt{iVoteID} and \texttt{PIN} from the voter, and subsequently modify their ballot, with a very low chance of detection by either the voter or the iVote server itself.

Furthermore, with Incapsula's cookies already being used to identify voters between both the registration server and voting server, it would also be trivial for such an attacker to link voters with their vote, removing the secrecy of their ballot and opening voters to the risk of vote-buying or coercion.

The device fingerprinting behaviour of the injected JavaScript may also allow these attacks to be performed in a selective fashion. Recent research by Cao {\it et al.} \cite{cao_fingerprinting} has shown that these fingerprinting methods can be used to identify users with a high degree of confidence, even across different browsers on the same device. This may provide an attacker with the ability to selectively target individual voters or electoral divisions, and to avoid targeting voters who may notice changes to the injected JavaScript (such as security researchers).

\subsubsection{Proof of Concept.}
%One might wonder how hard it would be to develop a malicious script to be injected that could compromise the voter credentials; the answer is surprisingly easily. 
We developed a short script that would leak the \texttt{iVoteID} and \texttt{PIN} by setting it in the profiling cookie. As such, the information would be leaked without need for any additional requests, making detection extremely difficult. Furthermore, due to the original injected script from Incapsula not being minimised, we were able to construct a malicious injection script that maintained all the functionality of the original, along with our additional malicious code, while still maintaining exactly the same length. 

%\code{lang=JavaScript, caption=Example of malicious code to include VoterNumber and PIN in profile cookie, label=lst:malicious_code}{code/inject_script_commented.js}

To achieve this we added two {\sf onChange} listeners to the \texttt{iVoteID} and \texttt{PIN} input boxes. We use these {\sf onChange} listeners to take a copy of the values entered and set them inside the profiling cookie. The advantage of this is that we are not adding any additional cookies, or requests, in order to leak the information, but instead using an existing side channel. 

In order to facilitate this we had to extend the lifetime of the profiling cookie. During testing we extended it to 1 hour, but realistically it only needs to be extended by a few minutes, the only requirement is that the cookie exists at the point the \texttt{iVoteID} and \texttt{PIN} is entered by the voter.

%An alternative approach, which would have been somewhat simpler, would have been to just set an additional cookie, however, we wanted to minimise the changes to demonstrate how the attack could be performed in a manner that is extremely difficult to detect. Additional cookies or requests would be easier to trace than detecting additional content in an existing cookie. 

%\code{lang=JavaScript, caption=Contents of Cookie (some content removed for brevity), label=lst:cookie}{code/cookie.js}

%The contents of the modified cookie, shortened for brevity, is \texttt{navigator=true, navigator.vendor=, ... ,digest=\#12345678,123456}, with the last two values, ``12345678" and ``123456" being the demo \texttt{iVoteID} and \texttt{PIN}.

\subsection{Foreign Access to TLS Private Keys}

In this attack scenario a cloud provider uses the brute force attack described in Section~\ref{sec:bruteForce} to recover the {\tt iVoteID} and {\tt PIN} from the passively observed {\tt voterID} value sent by the browser at login time. In comparison to the script injection attack above, this approach is completely passive and has the benefit of being undetectable at the cost of increased computational resources. Any cloud provider, therefore, must be trusted not to pursue such an attack unless the combined ID/PIN space was made cryptographically strong.

A more interesting scenario is one in which the cloud provider (a multinational company operating in many jurisdictions) must inadvertantly grant a foreign power the ability to man-in-the-middle an election through the course of prosecuting an otherwise lawful national security request.

As discussed in Section~\ref{international-cert-footprint}, valid TLS certificates for \url{*.elections.wa.gov.au} are served by Incapsula servers worldwide, with the associated TLS private keys also stored on these servers. The TLS certificates served by Incapsula's servers are multi-use certificates covering a number of domains, as described in Section~\ref{large-scale-cert-sharing}. This design has significant implications for the security of the TLS private keys associated with these certificates. 

%Access to these TLS private keys would provide an attacker with the ability to perform man-in-the-middle attacks on the TLS connection between voters and Incapsula's servers. %commented out as is similar to conclusion paragraphy

For example: a foreign government, as part of a legitimate domestic surveillance operation, may request that Incapsula provide access to the TLS private key for the domain \url{*.example.com} served by a PoP located in the foreign country. If this domain is contained in the same TLS certificate as \url{*.elections.wa.gov.au}, obtaining this private key would also provide the foreign government with the ability to perform man-in-the-middle attacks on voters using iVote. 
%!TEX root = ../main.tex

\section{Additional Findings}
\label{sec:additional}

\subsection{Verifiability} \label{sec:verifiability}

The iVote system incorporates a telephone verification service \cite{wa_how_to_ivote}, which allows a voter to dial a provided number and connect with an interactive voice response (IVR) system.

The telephone verification service requires the voter's \texttt{iVoteID}, \texttt{PIN}, and the receipt number provided by the iVote server after a vote has been successfully cast. After these three numbers have been provided, the telephone verification service reads back the list of candidates, in preference order, chosen by the voter in their completed ballot.

During the 2015 New South Wales state election, which also used the iVote system, Halderman and Teague identified several potential attacks against this telephone verification system \cite{halderman2015new}. These attacks could allow an attacker who had manipulated iVote ballots to avoid detection by voters who were attempting to verify that their vote was cast as intended.

One of these attacks is known as a ``clash attack," and is designed to trick voters by manipulating the registration and vote confirmation pages to provide the \texttt{iVoteID}, \texttt{PIN}, and receipt number of a previous like-minded voter with the same candidate preferences. The previous voter's ballot has been allowed to be recorded unmodified, and is then used as verification evidence for multiple voters. The actual votes of these voters can then be manipulated at-will with little chance of detection.

Crucially, the clash attack relies on accurate prediction of how a voter will vote prior to registration, so that they can be provided with the \texttt{iVoteID} and \texttt{PIN} of a like-minded voter who has submitted an unmodified ballot. In addition, the attack relies upon providing voters with a PIN rather than allowing them to choose one. This may raise the suspicions of voters who are aware that the iVote system is supposed to allow them to choose their own PIN.

For the 2017 WA State Election, the clash attack could be significantly improved as a consequence of Incapsula being used to proxy all voter connections to both the registration and voting servers. An attacker with access to Incapsula's systems could directly link each voter's registration details with their completed ballot, provided that the voter registers and votes using the same browser (and potentially across browsers as well \cite{cao_fingerprinting}).

Due to Incapsula's position as a DDoS mitigation service for a number of other online services, such an attacker would also have the ability to identify voters (and their likely voting preferences) with significantly more accuracy than if they only had access to the iVote system itself. This would allow for more accurate clash attacks to be performed.

\subsection{Bypassing DDoS Mitigation} \label{sec:bypassingDDoS}

It is assumed that the use of Incapsula's service to proxy iVote connections was an attempt to protect the iVote system from potential Distributed Denial of Service (DDoS) attacks during the 2017 WA state election.

DDoS mitigation services such as Incapsula operate by intercepting connections to a service (in this case, iVote), thereby hiding the true public IP Address of the service. If this protection is applied correctly, any attacker wishing to attack the iVote system will be forced to do so via Incapsula's systems---thereby allowing Incapsula's robust infrastructure to withstand the attack and filter legitimate connections through to the iVote system. For this protection to be effective, the true IP address of the service must be properly hidden from attackers \cite{vissers2015maneuvering}.

During the first several days of voting in the 2017 WA State Election, it was possible to identify the public IP address of the server hosting the iVote Core Voting System (CVS) for the WA election (\url{https://ivote-cvs.elections.wa.gov.au}), through specific requests to known iVote infrastructure in Sydney, NSW. This infrastructure could be publicly identified through DNS queries and other methods requiring little sophistication on the part of an attacker. With knowledge of this address, it would have been possible for an attacker to perform DDoS attacks against the iVote system directly, rendering Incapsula's protection ineffective.

Recommended practice for the use of DDoS mitigation services such as Incapsula is to prevent the identification of the true IP address of the service being protected, through techniques such as blocking all traffic from sources other than Incapsula itself \cite{incapsula_ddos_attacks,cloudflareDdosAttacks}. These protections were not correctly implemented  for the WA state election until we noticed the problem, several days after the opening of iVote,  and notified the WAEC. 

%!TEX root = ../main.tex

\section{Conclusion}
\label{sec:conc}
We have shown that utilizing cloud based DDoS protection servers can have a significant impact on the trust model of an internet based election. Furthermore, we have analysed the increased risks of tracking and interception associated with such services, and provided a proof of concept demonstrating how malicious JavaScript could be injected into a voting client in order to read or alter completed ballots. 

At the time of writing, more than two months after the election, the Western Australian Electoral Commission has published neither the raw voting data for iVote, nor the verification success and failure statistics.  Even if the votes were broadly similar to those cast on paper, and the verification failure rate was small, that would not constitute genuine evidence that the votes were accurately recorded. A lack of transparency in the process is simply no longer acceptable. In light of the trusted nature of cloud providers, their single point of failure, and the remote nature of potential attackers, the need for evidence-based election outcomes %in online voting systems 
is greater than ever.

\section*{Acknowledgements}

The authors thank the Western Australian Election Commission for quick acknowledgement and response to our disclosure. Thanks also to Yuval Yarom and the anonymous reviewers for helpful feedback.

\bibliographystyle{splncs03}
\bibliography{references}

\end{document}